\documentclass[preprint,superscriptaddress]{revtex4-1}

\usepackage{graphicx}
\usepackage{latexsym}
\usepackage{amsmath}
\usepackage{amssymb}
\usepackage{amsfonts}
\usepackage{color}
\usepackage{bm}
\usepackage{verbatim}
\usepackage{pagecolor,lipsum}
\usepackage{float}
\usepackage{comment}
\usepackage{siunitx}

    \makeatletter
\def\@fnsymbol#1{\ensuremath{\ifcase#1\or \dagger\or *\or \ddagger\or
   \mathsection\or \mathparagraph\or \|\or **\or \dagger\dagger
   \or \ddagger\ddagger \else\@ctrerr\fi}}
    \makeatother

\begin{document}

\title{Universal size-dependent nonlinear charge transport in single crystals of the Mott insulator Ca$_2$RuO$_4$}

\date{\today}

\author{G. Avallone}
\thanks{Authors contributed equally}
\affiliation{Huygens-Kamerlingh Onnes Laboratory, Leiden University, P.O. Box 9504, 2300 RA Leiden, The Netherlands.}
\affiliation{Dipartimento di Fisica “E.R. Caianiello”, Università degli Studi di Salerno - Via Giovanni Paolo II, 132 - I-84084 - Fisciano (Sa), Italy.}
\author{R. Fermin$^{\text{*}}$}
\thanks{Authors contributed equally}
\affiliation{Huygens-Kamerlingh Onnes Laboratory, Leiden University, P.O. Box 9504, 2300 RA Leiden, The Netherlands.}
\author{K. Lahabi}
\affiliation{Huygens-Kamerlingh Onnes Laboratory, Leiden University, P.O. Box 9504, 2300 RA Leiden, The Netherlands.}
\author{V. Granata}
\affiliation{Dipartimento di Fisica “E.R. Caianiello”, Università degli Studi di Salerno - Via Giovanni Paolo II, 132 - I-84084 - Fisciano (Sa), Italy.}
\author{R. Fittipaldi}
\affiliation{CNR-SPIN, c/o Università degli Studi di Salerno - Via Giovanni Paolo II, 132 - I-84084 - Fisciano (Sa), Italy.}
\author{C. Cirillo}
\affiliation{CNR-SPIN, c/o Università degli Studi di Salerno - Via Giovanni Paolo II, 132 - I-84084 - Fisciano (Sa), Italy.}
\author{C. Attanasio}
\affiliation{Dipartimento di Fisica “E.R. Caianiello”, Università degli Studi di Salerno - Via Giovanni Paolo II, 132 - I-84084 - Fisciano (Sa), Italy.}
\author{A. Vecchione}
\affiliation{CNR-SPIN, c/o Università degli Studi di Salerno - Via Giovanni Paolo II, 132 - I-84084 - Fisciano (Sa), Italy.}
\author{J. Aarts}
\thanks{Corresponding author}
\email{fermin@physics.leidenuniv.nl, aarts@physics.leidenuniv.nl}
\affiliation{Huygens-Kamerlingh Onnes Laboratory, Leiden University, P.O. Box 9504, 2300 RA Leiden, The Netherlands.}

\begin{abstract}
The surprisingly low current density required for inducing the insulator to metal transition has made Ca$_2$RuO$_4$ an attractive candidate material for developing Mott-based electronics devices. The mechanism driving the resistive switching, however, remains a controversial topic in the field of strongly correlated electron systems. Here we probe an uncovered region of phase space by studying high-purity Ca$_2$RuO$_4$ single crystals, using the sample size as principal tuning parameter. Upon reducing the crystal size, we find a four orders of magnitude increase in the current density required for driving Ca$_2$RuO$_4$ out of the insulating state into a non-equilibrium (also called metastable) phase which is the precursor to the fully metallic phase. By integrating a microscopic platinum thermometer and performing thermal simulations, we gain insight into the local temperature during simultaneous application of current and establish that the size dependence is not a result of Joule heating. The findings suggest an inhomogeneous current distribution in the nominally homogeneous crystal. Our study calls for a reexamination of the interplay between sample size, charge current, and temperature in driving Ca$_2$RuO$_4$ towards the Mott insulator to metal transition.
\end{abstract}

 \maketitle

\section{Introduction}

Strongly correlated electron systems, and in particular the metallic oxides, have truly marked a paradigm shift in physics, inspiring a decades-long search for exciting quantum materials to explore exotic phenomena, such as high-temperature superconductivity,\cite{Keimer2015} electron hydrodynamics,\cite{Moll2016,Sulpizio2019} and holography.\cite{Maldacena1997,Hartnoll2016} A prime example of this, is the family of ruthenates witch exhibits a rich and diverse phase diagram that includes Mott insulators,\cite{Nakatsuji1997,Zhu2016} unconventional superconductivity,\cite{Maeno2012,Leggett2020} and magnetism.\cite{Cao1997,Grigera329} The 4\textit{d} electron Mott insulator Ca$_2$RuO$_4$\cite{Nakatsuji1997} (hereafter Ca214) has become the subject of intense research in recent years due to its intriguing electrical~\cite{Alireza2010,Nakamura2002,Cirillo2019,Bertinshaw2019,Fursich2019_Raman,Zhang2019,Alexander1999,Nakatsuji1999,Nakatsuji2000a,NAKATSUJI2001,Sutter2019,Nakamura2002,Steffens2005,Nakamura2007} and magnetic properties.\cite{Braden1998,Zegkinoglou2005} Specifically, at room temperature it undergoes a current-driven insulator to metal transition (IMT) which occurs at unusually low electric ($E$-)field or current density ($J$) thresholds ($\sim 40$ V/cm or few A/cm$^2$ respectively).\cite{Nakamura2013,Okazaki2013} This is in contrast to previous reports on Mott insulators, where the IMT is limited to low temperatures and/or to the application of high $E$-fields.\cite{Taguchi2000,Kanki2012} The capacity to switch between resistive states at room temperature is a desirable property to realize current switchable memories, neuromorphic devices, and next-generation oxide electronics.\cite{Yang2011,stoliar2017}

The underlying mechanism responsible for such low current densities is still a topic of intense debate. Due to its high resistivity in the insulating state, Joule heating plays a significant role since the IMT can also be thermally driven by heating the crystal above 357 K.\cite{Alexander1999} This has led to ambiguity about the origin of the IMT, and whether it is thermally or electronically driven, as evident by the large number of works discussing local heating.\cite{Okazaki2013,Mattoni2020,Jenni2020,Okazaki2020,Terasaki2020,Millis2020,Zhang2019} Specifically, in a recent work, it was found that the IMT is always accompanied by a local temperature increase to the transition temperature.\cite{Mattoni2020} Moreover, as a precursor to the metallic phase, a third and non-equilibrium (also called metastable) phase has been detected by X-ray and neutron diffraction experiments, and additionally with Raman spectroscopy.\cite{Cirillo2019,Bertinshaw2019,Fursich2019_Raman} This non-equilibrium phase seems to be induced by low current densities, but its role in the current-driven IMT remains elusive. Another major hurdle is the pronounced structural transition accompanying the IMT, in which the RuO$_2$ octahedra are elongated, and the unit cell is transformed from orthorhombic to tetragonal.\cite{Friedt2001,Gorelov2010,Nakamura2002,Nakamura2013,Cirillo2019,Bertinshaw2019,Zhang2019} This transition leads to a strong temperature dependence of the unit cell volume ($\sim$1\% between 100 K and 400 K), introducing large internal strains in the crystal, often resulting in the formation of cracks or even shattering the crystal upon reentering the insulating phase.\cite{Friedt2001}

Although it is much debated whether the IMT is primarily triggered by Joule heating or driven by electronic effects, a parameter that has been left unexplored in this discussion is the \textit{size} of the samples. Decreasing the size of bulk samples down to \SI{}{\micro\metre} range gives larger control over current paths in the crystal (due to the uniform rectangular cross section). Furthermore, micro cracks and step-like terraced edges, which occur naturally in mm-sized bulk crystals, are scarce in microscopic samples. Also, since the voltage contacts cover the full side of the microscopic samples, current injection is more homogeneous. This is in contrast with mm-sized crystals where point contacts, and their entailing current crowding effects, are more common. Finally, due to the enhancement of the surface-to-volume ratio and the direct contact with an isothermal substrate, the microscopic samples are expected to be less susceptible to thermal gradients and heating effects.

Here we carry out an extensive size-dependent study using a large number (39) of ultra-pure single crystal samples of Ca214, ranging between the hundreds of nm and the millimeter scale. We examine the role of the current density in inducing the non-equilibrium phase, which we can probe reversibly (i.e., without inducing the IMT and consequently damaging or altering the crystal). We find a surprising enhancement of the required current density by at least four orders of magnitude upon decreasing the cross section. By integrating a micrometer-sized platinum thermometer, we are able to directly probe the local temperature of the microscopic samples, and demonstrate that the pronounced size dependence is not caused by thermal effects. Our findings call for a careful reexamination of the relevant mechanism behind the non-equilibrium phase and its relation to the IMT in Ca214.

\section{Results}

\subsection{Characterizing microscopic samples} 

\begin{figure*}[bt!]
 \centerline{$
 \begin{array}{c}
  \includegraphics[width=1\linewidth]{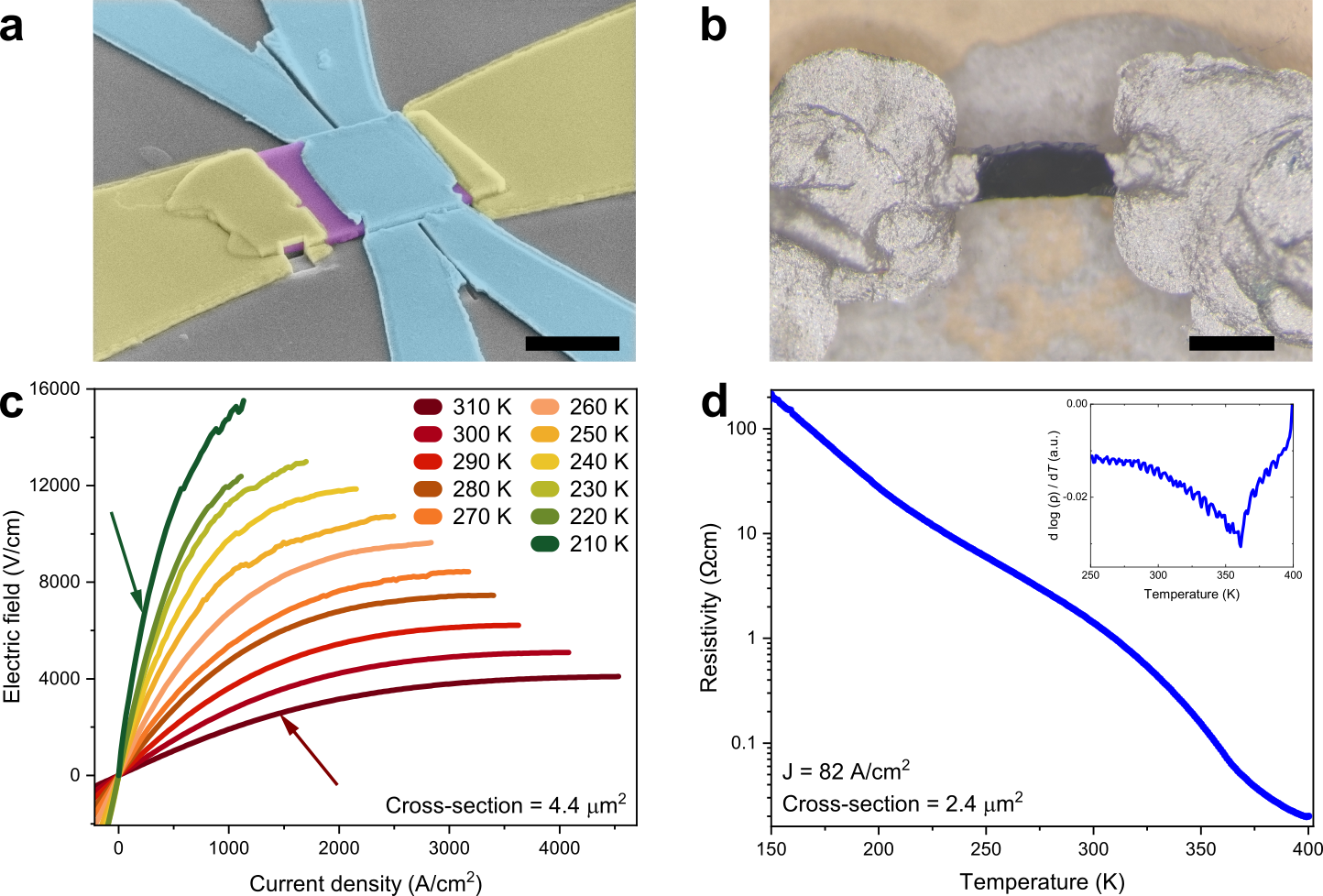}
 \end{array}$}
 \caption{\textbf{Transport characteristics of microscopic samples.} \textbf{a} Displays a false colored scanning electron micrograph of a microscopic sample (in purple) that is contacted by yellow colored Ti/Au contacts. In blue the embedded Pt-thermometer, that can be employed to locally measure the temperature, is shown. The scale bar represents \SI{3}{\micro\metre}. \textbf{b} Shows an optical microscope image of a mm-sized sample, where the scale bar measures \SI{500}{\micro\metre}. Panel \textbf{c} shows $E(J)$-characteristics acquired on a microscopic sample at different temperatures. The arrows indicate the current density that would correspond to the Figure of Merit. Note that these curves are obtained below the IMT. The resistivity versus temperature of a typical microscopic sample is shown in panel \textbf{d}. In the microscopic size range we do not observe a sharp temperature induced IMT, however, as indicated by the inset (the derivative of the curve in the main figure), there is a clear inflection point indicating the thermal transition.} \label{figure1}
\end{figure*} 

To examine the crossover from large mm-sized bulk behavior to the microscopic one, we have produced samples of varying cross section between \SI{0.5}{\milli\metre^2} and \SI{0.5}{\micro\metre^2}. The microscopic samples were fabricated using mechanical exfoliation.\cite{ThinFlakeProduc} This enables us to produce micron-sized "crystal flakes", which can be lithographically contacted for electrical transport measurements, without compromising the material quality. In addition to the microscopic samples, we have fabricated mm-sized bulk samples that are hand-contacted by the use of silver paint. We employed a focused Ga$^{+}$-ion beam (FIB) as the principal structuring technique throughout this work. This provides us with the means to control the cross section in a systematic manner. Figures 1a and 1b show scanning electron microscope and optical microscope images of samples in microscopic and mm-sized ranges respectively.

All measurements are carried out by biasing a current in the ab-plane rather than a voltage, since inducing the metallic state by the application of electric field is more abrupt and therefore the crystals are more likely to break. Furthermore, current density as a function of electric field is hysteretic which complicates the study of size effects carried out here. We have conducted electric field versus current density ($E(J)$-characteristic) and resistivity measurements as a function of temperature and found that the microscopic samples qualitatively show similar behavior as the mm-sized bulk samples.\cite{Cirillo2019} Typical results from microscopic samples are summarized in Figures 1c and 1d. It is important to note that, despite several orders of magnitude of size variation, we find the room temperature resistivity of all our samples (microscopic and mm-sized bulk) to be comparable, and in good agreement with the literature values.\cite{Nakatsuji1997} Curiously, however, the temperature-induced IMT is broadened in microscopic samples with respect to those observed in mm-sized bulk samples (see Figure 1d). We attribute the broadening of the transition to the small sample thickness since similar effects are observed in thin film samples.\cite{Tsurumaki-Fukuchi2020,Miao2012,Wang2004,Dietl2018} We did observe an abrupt change in resistivity upon cooling a relatively thick (T = \SI{1.5}{\micro\metre}) microscopic sample through the metal to insulator transition for further discussion on this aspect the reader is referred to the Supplementary Note 2.

\subsection{Comparing samples of different cross section}\label{size_study}

Figure 1c shows $E(J)$-characteristics obtained on a microscopic sample at different temperatures. These curves qualitatively show similar behavior as the mm-sized bulk samples. However, the current density required for triggering the metastable phase, which is indicated by a negative $\text{d}E/\text{d}J$ as a function of $J$, exceeds the literature values (for mm-sized crystals) by at least four orders of magnitude.\cite{Zhang2019,Nakamura2013,Mattoni2020} This is demonstrated in Figure 2a, where we plot the differential resistivity $\text{d}E/\text{d}J$ as a function of $J$ for selected samples. The current density at which $\text{d}E/\text{d}J$ begins to decline, corresponding to nonlinear conduction, increases if the sample size is reduced to microscopic scales.
In order to compare the nonlinear conduction under applied current between different samples in a systematic manner, we propose a Figure of Merit (FOM) that can be applied to samples of different sizes (see Figure 2). First, differential resistivity versus current density curves are computed by analytically differentiating the $E(J)$-characteristics (Figure 2a). The curves are then normalized with respect to their low current value (i.e., currents for which the resistivity is current independent and conduction is linear). We choose the current density at which the slope of the $E(J)$-curve has halved as our Figure of Merit, called the 50\%-slope current density. The choice for this FOM is suitable since we can compare the size dependence of the $E(J)$-characteristic in a regime where heating effects play a relatively unimportant role. Furthermore, at the FOM current density, neither the metallic phase nor even the metastable phase is induced in the sample.\cite{Cirillo2019} However, we find the shape of the $E(J)$-curves to be the same between all measured samples. Therefore, the current density required for inducing the metastable phase scales with the FOM current density accordingly, which allows for determining the current density needed to induce the metastable phase. Lastly, the slope of the differential resistivity appears to be maximal at the FOM. Therefore, the choice of this FOM leads to a small uncertainty in the estimated current density, which would not be the case when using the current density at the maximum $E$-field, for instance.

\begin{figure*}[bt!]
 \centerline{$
 \begin{array}{c}
  \includegraphics[width=1\linewidth]{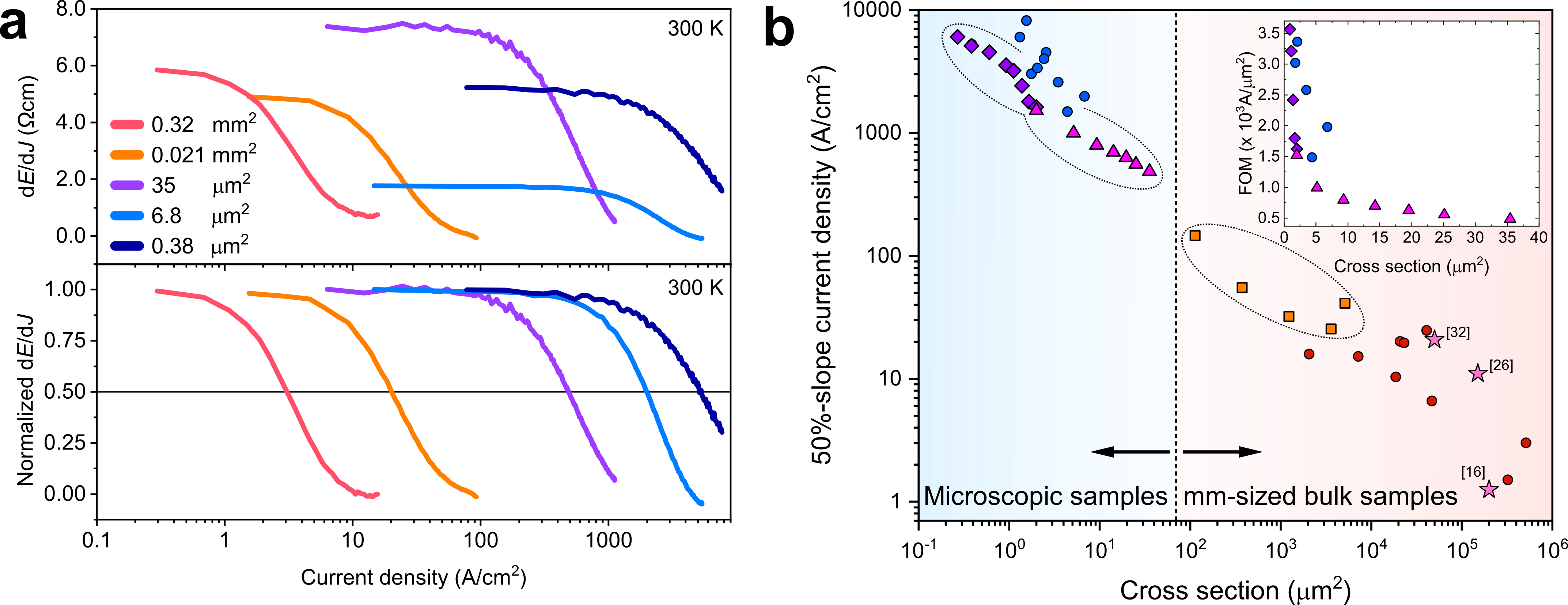}
 \end{array}$}
 \caption{\textbf{Size dependence of nonlinear conduction.} The top panel of \textbf{a} shows the differential resistivity $\text{d}E/\text{d}J$ as a function of current density of a few selected samples at room temperature. The legend shows the cross-sectional area of these samples. For all microscopic and mm-sized bulk samples the low current resistivity at room temperature is found to be corresponding to the literature values. In the lower panel of \textbf{a}, the differential resistivity of these samples is normalized with respect to the low current resistivity (i.e., where the resistivity is current independent). Even though the $E(J)$-characteristics of all samples are qualitatively the same, the current density scale at which they exhibit nonlinear conduction differs orders of magnitude. The current density at which the slope of the $E(J)$-characteristic is halved, is chosen as the Figure of Merit (FOM), which corresponds to the horizontal reference line. \textbf{b} shows the FOM as a function of cross-sectional area for all measured samples at room temperature on a log-log scale. N.B. each circular point corresponds to a single sample. The thinning study samples are depicted with a unique non-circular symbol and a different color. The inset displays the FOM vs sample cross section for selected samples on linear scales. The star-symbols correspond to the FOM extracted from literature; the label indicates the reference number.} \label{figure2}
\end{figure*} 

Figure 2b displays the FOM as a function of the sample cross section for all measured samples. The current density at which nonlinear conduction sets in, grows monotonically with decreasing sample dimensions and, curiously, the FOM shows a power law dependence on the cross-sectional area. To exclude any potential artifacts associated with sample preparation, we performed a thinning study on three samples over a large size range. We incrementally modify the width and thickness of the same crystal through consecutive FIB structuring steps, and measure the $E(J)$-characteristic, after each one. Following this procedure, we changed the width of a single microscopic sample in up to seven different thinning steps. Moreover, in an mm-sized bulk sample, we were able to decrease the sample width by a factor 50. Samples on which a thinning experiment was carried out are encircled and highlighted in Figure 2b by a non-circular symbol and a different color. The FOM as a function of only the width or thickness of the sample (see Supplementary Note 3) reveals the size dependence is not exclusively formed by any of these two parameters but is rather a combination between them, which means that neither of these dimensions is more significant in determining the transport properties of the crystal. The simplest combination of the two parameters is the cross section, which is chosen in Figure 2b.  Supplementary Figure 5 shows the FOM as a function of volume.

Despite the clear trend in Figure 2b, there is a sample to sample variation that cannot be explained by the uncertainty in the measured crystal dimensions. The deviation from the trend in the mm-sized samples can be explained by irregular current paths in the sample and inconstant cross-sectional area throughout the length of the sample. For the microscopic samples, this does not hold. However, the ratio between width and thickness differs among the samples, and therefore they are expected to respond differently to decreasing dimensions if surface layer effects are important. Furthermore, oxygen relocation can play a role in modifying the crystal structure when passing a current, resulting in a change of transport properties between current cycles. This is consistent with our observations, where, in some cases, the $E(J)$-characteristic slightly changes after different current or temperature cycles. These arguments can explain the increase in spread observed in Figure 2b.

\subsection{The role of temperature}\label{temperature}

All the experiments reported previously, where the temperature is locally measured in Ca214, are carried out on mm-sized bulk samples using optical techniques, as local contact thermometry is experimentally difficult to realize at these dimensions. Microscopic samples do not have this limit and enable contact thermometry to give insight into the local sample temperature. To implement this, we carefully designed and fabricated a platinum thermometer on top of selected microscopic samples (see Figure 1a). Since platinum has a distinctly linear temperature dependence, we are able to accurately measure the temperature of the microscopic samples, while simultaneously driving a current using an independent bias (see Supplementary Note 1 for details).

In order to make plausible that the temperature difference between the Pt and the sample is negligible and to investigate the likelihood of potential temperature gradients in the sample, we have performed thermal simulations accompanying the thermometry results. Here we solved for a steady-state temperature, under the assumption that heat is generated uniformly in the crystal flake while the substrate temperature remains constant \SI{50}{\micro\metre} away from the sample. The full technical details of these simulations can be found in Supplementary Note 6.

 \begin{figure}[htb!]
 \centerline{$
 \begin{array}{c}
  \includegraphics[width=1\linewidth]{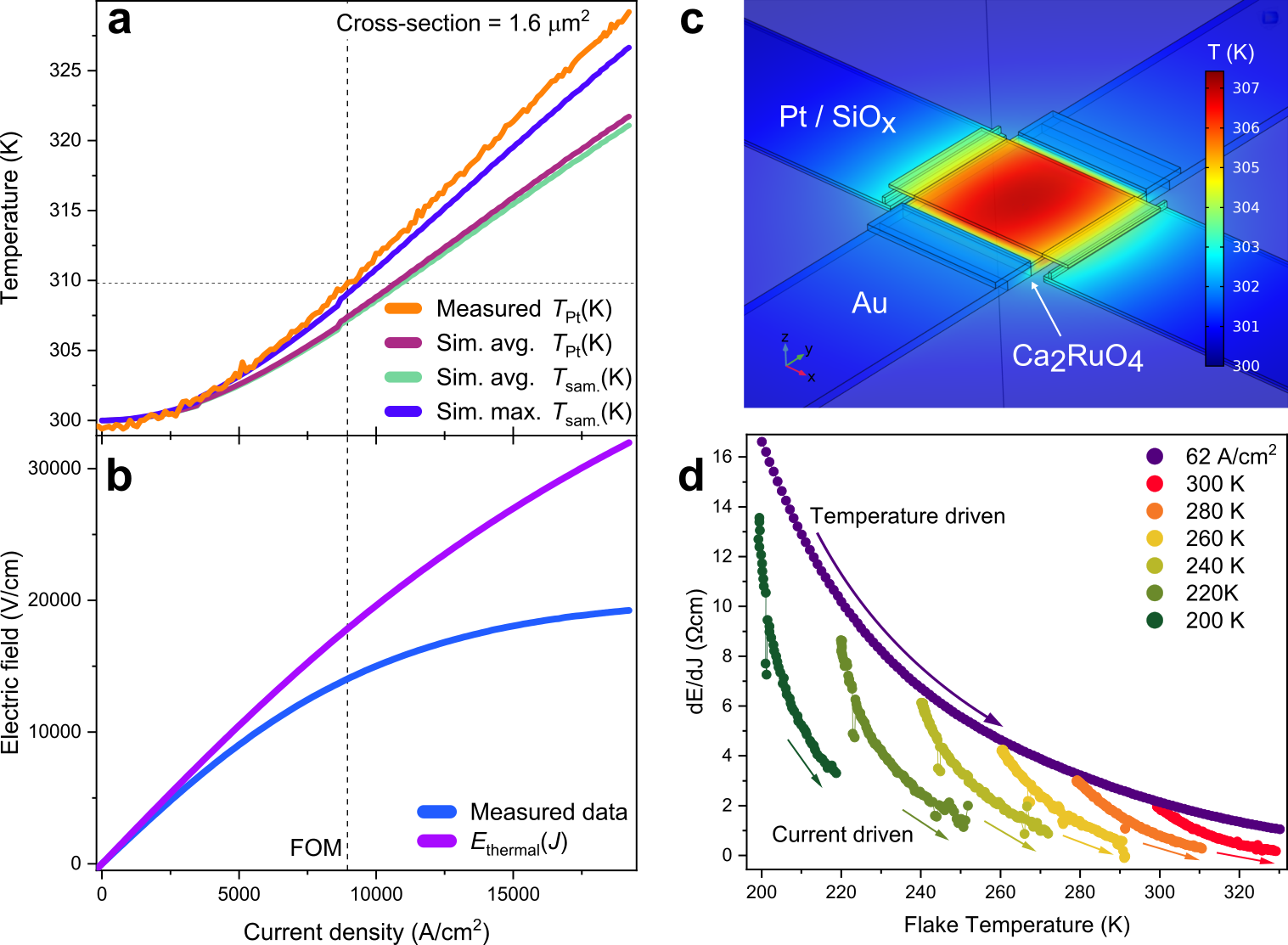}
 \end{array}$}
 \caption{\textbf{The role of temperature in microscopic samples.} \textbf{a} displays the local measured Pt temperature and multiple simulated temperatures as a function of the applied $J$. The data is aquired on the sample depicted in Figure \ref{figure1}a. The substrate temperature is 300 K. \textbf{b} shows the electric field acquired simultaneously with the temperature measurement. The measured data is compared to $E_\mathrm{thermal}$ (i.e., the calculated electric field on basis of Joule heating only). The vertical reference line indicates the FOM. The local temperature increase at the FOM is below 10 K. A simulated temperature heatmap of the sample at the FOM current density is shown in \textbf{c}. \textbf{d} displays the differential resistivity as a function of the flake temperature. Either the background temperature or the current density is changed. At all current densities, the nonlinear conduction is stronger than heating effects can provide.}\label{figure3}
 \end{figure}

Figure 3a displays the platinum temperature obtained on the sample shown in Figure 1a (orange curve) together with the simulation results as a function of current density. The measured temperature is exceeding the expected averaged Pt temperature (purple curve) for high currents. This can be caused by the absence of heat transfer barriers between the simulated elements, whereas in experiments these can be expected. The purple curve can therefore be considered as a lower bound of the expected Pt temperature. Nevertheless, the simulated average sample (green curve) and Pt temperature (purple curve), show correspondence to a high degree. We have also examined temperature gradients in the sample and found temperature differences to be limited to 3 K in the sample (see Figure 3c and Supplementary Note 6). Furthermore, we evaluate the maximum temperature within the simulated flake geometry (blue curve) and compare it to the simulated average Pt temperature. We find these to differ less than 2 K at the FOM current density. Therefore, we consider our Pt thermometry technique suited for acquiring the average sample temperature at the FOM. At the FOM current density, the locally measured temperature increase is found to be less than 10 K. Therefore, we conclude that Joule heating is non-zero but similar to mm-sized bulk samples.\cite{Mattoni2020}

The measurement of the local temperature enables us to calculate what the $E(J)$-characteristic would look like based on heating effects only. For the resistivity versus temperature measurement on this sample, we compute the d\textit{E}/d\textit{J} from our $E(J)$-characteristics for low currents (i.e., where the resistivity $\rho$ is current independent):

\begin{align} \label{Eq:1}
\rho(T) = \frac{\mathrm{d}E}{\mathrm{d}J}(T, J \approx 0)
\end{align}

\noindent Since the temperature is homogeneous up to a variation of 3 K over the entire sample, we can use the locally measured temperature as a function of current density to find the d\textit{E}/d\textit{J} as a function of heating effects:

\begin{align} \label{Eq:2}
\frac{\mathrm{d}E_\mathrm{thermal}}{\mathrm{d}J}(J) = \frac{\mathrm{d}E}{\mathrm{d}J}(T(J), J \approx 0) = \rho(T(J))
\end{align}

\noindent Integrating the equation \eqref{Eq:2}, we can reconstruct the $E(J)$-characteristic that would have been measured if Joule heating was the only mechanism causing the nonlinear conduction. We set the integration constant by requiring zero electric field for zero current density ($E$(0) = 0; see Supplementary Note 5 for more details):

\begin{align} \label{Eq:3}
E_\mathrm{thermal}(J) = \int \frac{\mathrm{d}E_\mathrm{thermal}}{\mathrm{d}J}(J) \mathrm{d}J
\end{align}

\noindent We denote this reconstructed function as $E_\mathrm{thermal}(J)$ and plot it alongside the measured data in Figure 3b. Even at low current densities, the actual measured data shows stronger nonlinear conduction than $E_\mathrm{thermal}(J)$, indicating that current-driven effects are significant in the microscopic samples for all applied current densities probed in this study.

 Alternatively, the differential resistivity as a function of local sample temperature is plotted in Figure 3d. The purple curve describes the resistivity as a function of temperature with a constant current density (62 A/cm$^2$). Alongside, we plot the differential resistivity as a function of increasing bias current, while maintaining a constant substrate temperature. Evidently, the slope decrease of $E(J)$-characteristic, caused by the application of current, is not primarily driven by Joule heating.

We conclude based on our thermometry experiments that the size dependence presented in Figure 2b cannot be explained by Joule heating. Furthermore, we find that a current-driven mechanism is present parallel to Joule heating in the non-equilibrium phase.

\section{Discussion}

Before discussing the possible origin of the observed size dependence, we consider the following notes. Firstly, there seems to be no intrinsic length scale found in our measurements. This can be seen from the absence of a specific cross section where the FOM changes discontinuously. Secondly, we can exclude the effects of microcracks since the size dependence continues even in the microscopic regime where micro cracks are not present. By inspecting the samples with a scanning electron microscope, we could confirm that no micro cracks were induced by measuring the $E(J)$-characteristic. Thirdly, the observation of a current-driven mechanism parallel to Joule heating, as observed here, was recently also reported by Jenni et al.\cite{Jenni2020} Lastly, we note that our findings are consistent with the reports on epitaxially grown thin films of Ca214, where the current density required to induce the IMT was also found to be many orders of magnitude higher than in bulk literature values.\cite{Tsurumaki-Fukuchi2020}

The emergence of the metallic phase was recently attributed by Terasaki et al.\cite{Terasaki2020} to energy flow, as opposed to charge transport. In this study, the energy flow corresponds to the dissipated power (product of current and voltage $IV$), in contrast, here it is more insightful to describe the product of electric field and current density $EJ$ (power dissipation density), enabling the comparison of samples of different length scales. Since microscopic samples show a four order of magnitude increase in current density at which we observe nonlinear transport, we find that the power dissipation density is eight orders of magnitude larger in microscopic samples than in mm-sized bulk samples (see Supplementary Figure 8 a). It is therefore unlikely that the energy flux through the crystal is responsible for observed size dependence.

Alternatively, the size dependence might be explained by strain, induced through coupling to the substrate. In the zero-current limit, however, the substrate does not exercise any strain on the crystal, as we will discuss below. The room temperature resistivity of Ca214 is strongly dependent on strain and the IMT can be induced by applying ~0.5 GPa of pressure.\cite{Nakamura2002,Steffens2005,Nakamura2007} By using ultra-low currents ($\backsim$nA), we have confirmed that the insulating-state resistivity of the microscopic samples matches that of their mm-sized counterparts, signaling the absence of strain. Moreover, in a previous work on the isostructural unconventional superconductor Sr$_2$RuO$_4$, we have prepared microscopic samples using equal methods and substrates.\cite{Yasui2020} These samples retained bulk properties to sizes below 200 nm, whereas, like Ca214, its transport properties are heavily dependent on strain.\cite{Grinenko2021,Barber2018,Steppke2017} Therefore we conclude that our microscopic samples do not experience strain at zero current bias. In simultaneous transport and X-ray diffraction measurements, it has been shown that at the FOM current density, there is no detectable change of the lattice parameters.\cite{Cirillo2019} Combined with the absence of strain at zero current bias, this leads to the conclusion that our samples do not experience any strain at the FOM, regardless of their size. Thus, we do not regard strain as a plausible explanation for the observed size dependence.

Near the IMT, and independent of whether that is induced by current or temperature, the arguments supporting the absence of strain no longer hold, as the lattice constants significantly alter.\cite{Cirillo2019,Bertinshaw2019} Although we are far below the IMT in our current-driven experiments, we pass the IMT in the temperature sweep, presented in Figure 1d. Strain-related effects could be responsible for the observed broadening of the transition. Moreover, a broadened transition is commonly observed in thin films, where strain induced by the substrate is expected to be an influential parameter.\cite{Tsurumaki-Fukuchi2020,Miao2012,Wang2004}

Finally, an inhomogeneous distribution of current throughout the cross-sectional area could potentially explain the size dependence found here. Since the current density is calculated by dividing the applied current by the entire cross section, the apparent current density in mm-sized bulk samples might be lower than the actual current density that is physically relevant. Due to their size and geometry, however, in the microscopic samples, the apparent and actual current densities can be more similar.

Metallic filament formation, which is known to occur in Mott insulators,\cite{Lange2020} may induce such highly inhomogeneous current distributions. Contrarily, for Ca214, Zhang et al.\cite{Zhang2019} inspected phase separation using scanning near field optical microscopy (SNOM) and found a ripple pattern at the phase boundary between the insulating and metallic states. Although the resolution of the SNOM technique might be insufficient to rule out sub 100 nm channel formation, it is difficult to reconcile the striped pattern found in SNOM with filament formation at present.

Furthermore, Zhang et al. find that the metallic phase nucleates at the top of the sample, and its depth increases when traversing a phase boundary in the phase-separated state. This could suggest an inhomogeneous distribution of current throughout the cross-sectional area, that is high at the edges of the crystal and decreases towards the center of the bulk. Since the microscopic crystals have a larger surface-to-volume ratio, the edges are relatively more dominant. If the current density gradually increases below the current-carrying surface, this proposed mechanism will not feature any length scale at which a discontinuous change is expected in the FOM, which is compatible with our findings. In Supplementary Note S7, a minimal toy model describing the inhomogeneous current density is presented. This model can reproduce the power law dependence observed in Figure 2b.

In conclusion, we have performed a detailed study on size-dependent electrical properties of the Mott insulator Ca$_2$RuO$_4$. We find a surprising relation between crystal size and the current density at which nonlinear conduction occurs, which increases four orders of magnitude when the sample size is reduced from \SI{0.5}{\milli\metre^2} to \SI{0.5}{\micro\metre^2}. We have strong indications that the observed size dependence is not caused by Joule heating, using a local Pt-thermometer fabricated on top of selected microscopic samples. Our findings indicate an intrinsically inhomogeneous current density distribution in single crystals of Ca$_2$RuO$_4$. This calls for a reexamination of the relevant role of current in the metastable phase and its possible relation to the IMT. As an outlook, the combination of microscopic samples with a well-controlled current path, and local platinum temperature probes, provides a state-of-the-art approach to study the interplay between thermal and electronic effects, which can be used to study the IMT in Mott insulators.

\section{Methods}

\subsection{Crystal growth}

Bulk single crystals of Ca214 were grown by a flux feeding floating zone technique with Ru self-flux using a commercial image furnace equipped with double elliptical mirrors described elsewhere.\cite{Fukazawa2000,Granata2020} Several techniques, including X-ray diffraction, energy dispersive spectroscopy, and polarized light optical microscopy analysis, have been used to fully characterize the structure, quality, and purity of the crystals.

\subsection{Sample fabrication and transport experiments}

The microscopic samples were fabricated by the use of mechanical exfoliation on highly resistive SrTiO$_3$ or sapphire substrates. The flake samples are significantly thicker than monolayers. Therefore, random strain patterns, wrinkles, and folds associated with the thin film limit are absent. Due to the natural shape of the crystal flakes, the produced samples are suited best for passing current in the ab-plane. Therefore all experiments are carried out in this configuration.

Considerable effort was put into the fabrication of the electrical connections on the samples to reduce the contact resistance that possibly becomes an extrinsic origin for nonlinear conduction due to local Joule heating. To perform electrical transport measurements, the samples obtained by the exfoliation process were contacted by means of an electron beam lithography step followed by a Ti/Au sputtering deposition and a lift-off procedure. When necessary, we used electron beam induced deposition to provide an electrical reinforcement of the contacts by locally depositing an additional layer of tungsten-carbide on them. On the mm-sized samples, we found a resistive background in the $E(J)$-characteristics, from which we extracted the contact resistance to be less than 10 $\Omega$, which is significantly smaller than the sample resistance, supporting the use of a two-probe measurement. Some microscopic samples were contacted in a 4-probe geometry. Using these samples, we could confirm that the contact resistance is negligible in comparison to the sample resistance. In addition to the microscopic samples, we have fabricated conventionally mm-sized bulk samples that are contacted by hand using silver paint (Agar Scientific G3691).

The electrical transport measurements were carried out by applying a current using a Keithley 6221 current source. The voltage drop over the sample was measured using a Keithley 2182a nanovolt meter.

\subsection{Microscopic thermometry and FIB-processing}

In a second step of electron beam lithography, the Pt-thermometer was added to selected samples. The Pt-circuit is electrically isolated by a thin layer of sputter-deposited SiO$_\text{x}$ ($\sim$ 100 nm). This ensures that the Pt-thermometer and sample form two decoupled electrical systems. The resistance between the Pt-circuit and the Ca214 sample was found to be over 1 G$\Omega$ and the $E(J)$-characteristic measured before and after the application of the Pt-thermometer was found unaffected, as demonstrated in Supplementary Note 1.

We used the FIB technique to modify both the sample width and thickness enabling the full control of sample dimensions and geometry over a large crystal size range (see Supplementary Note 4). After fabrication, we have confirmed by scanning electron microscope inspection that the microscopic samples lack any micro cracks.

\section{Data availability statement}
The data that support the findings of this study are available from the corresponding
author upon reasonable request.

\section{Acknowledgements}
This work was supported by the Dutch Research Council (NWO) as part of the Frontiers of Nanoscience (NanoFront) program, and through NWO Projectruimte grant 680.91.128. This work benefited from access to the Netherlands Centre for Electron Nanoscopy (NeCEN) at Leiden University. The authors want to thank Jan Zaanen for discussing the results in an early stage of the work and Norman Bl\"umel for additional sample characterization.

\section{Competing interests}
The authors declare no competing financial or non-financial interests.

\section{Author contributions}
V. Granata, R. Fittipaldi and A. Vecchione grew the crystals and characterized them for purity. G. Avallone, R. Fermin and K. Lahabi prepared the microscopic samples and performed the measurements.  All authors contributed to writing the manuscript. G. Avallone and R. Fermin are considered co-first author.

\end{document}


\title{Supplementary information of: Universal size-dependent nonlinear charge transport in single crystals of the Mott insulator Ca$_2$RuO$_4$}

\date{\today}

\author{G. Avallone}
\thanks{Authors contributed equally}
\affiliation{Huygens-Kamerlingh Onnes Laboratory, Leiden University, P.O. Box 9504, 2300 RA Leiden, The Netherlands.}
\affiliation{Dipartimento di Fisica “E.R. Caianiello”, Università degli Studi di Salerno - Via Giovanni Paolo II, 132 - I-84084 - Fisciano (Sa), Italy.}
\author{R. Fermin$^{\text{*}}$}
\thanks{Authors contributed equally}
\affiliation{Huygens-Kamerlingh Onnes Laboratory, Leiden University, P.O. Box 9504, 2300 RA Leiden, The Netherlands.}
\author{K. Lahabi}
\affiliation{Huygens-Kamerlingh Onnes Laboratory, Leiden University, P.O. Box 9504, 2300 RA Leiden, The Netherlands.}
\author{V. Granata}
\affiliation{Dipartimento di Fisica “E.R. Caianiello”, Università degli Studi di Salerno - Via Giovanni Paolo II, 132 - I-84084 - Fisciano (Sa), Italy.}
\author{R. Fittipaldi}
\affiliation{CNR-SPIN, c/o Università degli Studi di Salerno - Via Giovanni Paolo II, 132 - I-84084 - Fisciano (Sa), Italy.}
\author{C. Cirillo}
\affiliation{CNR-SPIN, c/o Università degli Studi di Salerno - Via Giovanni Paolo II, 132 - I-84084 - Fisciano (Sa), Italy.}
\author{C. Attanasio}
\affiliation{Dipartimento di Fisica “E.R. Caianiello”, Università degli Studi di Salerno - Via Giovanni Paolo II, 132 - I-84084 - Fisciano (Sa), Italy.}
\author{A. Vecchione}
\affiliation{CNR-SPIN, c/o Università degli Studi di Salerno - Via Giovanni Paolo II, 132 - I-84084 - Fisciano (Sa), Italy.}
\author{J. Aarts}
\thanks{Corresponding author}
\email{fermin@physics.leidenuniv.nl, aarts@physics.leidenuniv.nl}
\affiliation{Huygens-Kamerlingh Onnes Laboratory, Leiden University, P.O. Box 9504, 2300 RA Leiden, The Netherlands.}

\pacs{} \maketitle

\section*{Supplementary notes}

\subsection{Measuring the local temperature of a microscopic sample with an embedded thermometer}

Measuring the local sample temperature proved to be vital in many proceeding studies.\cite{Okazaki2013,Mattoni2020,Jenni2020,Okazaki2020,Terasaki2020,Millis2020,Zhang2019} In order to have full insight on the local temperature, we fabricate a thermometer in direct contact with selected microscopic samples. After contacting the microscopic sample by Ti/Au contacts, using electron beam lithography and a lift-off procedure (see Supplementary Figure \ref{sub_fig_1}b), we sputter deposit a layer of approximately 100 nm SiO$_\text{x}$. In a second step of lithography, we design a patch over the microscopic sample that has four leads, enabling a 4-terminal measurement over the deposited material directly on top of the crystal, as depicted in Supplementary Figure \ref{sub_fig_1}d. The material of choice for the thermometer circuit is platinum since its resistivity is highly linear over the temperature range where we carry out our experiments, as can be seen in the thermometer calibration in Supplementary Figure \ref{sub_fig_1}c. The resistance of the Pt-thermometer changes by 0.27 K/m$\Omega$. Furthermore, the thickness of the SiO$_\text{x}$ layer was found to be optimal, providing both a disconnected electrical system from the crystal whereas remaining in thermal contact to provide sub-Kelvin measurement precision. Supplementary Figure \ref{sub_fig_1}a depicts the IV-characteristic acquired before and after the application of the thermometer. The induced change is found to be minimal as expected since the crystal and sample circuits are electrically decoupled. Generally, a resistance of over 1 G$\Omega$ is found between the crystal and the Pt-thermometer circuit.

Thermometry experiments were carried out on multiple microscopic samples. On thicker samples, it is technically more challenging to connect a Pt layer on top of the crystal, to the leads on the substrate. Therefore, these experiments are carried out on the smallest crystals only. To complement the data presented in the main text, we present a full data set obtained on another microscopic sample in Supplementary Figure \ref{sub_fig_7}.

\subsection{Abrupt transition in a relatively thick microscopic sample}

In most of the microscopic samples the temperature-driven IMT is broadened. Figure 1d of the main text shows the general resistivity versus temperature behavior of microscopic samples. However, on one of our microscopic samples with a relative large thickness (1.5 $\mu$m) with respect to the width (3 $\mu$m) we have acquired a resistivity versus temperature curve that is more resembling data acquired on a mm-sized bulk crystal (see Supplementary Figure \ref{sub_fig_2}a). We observe a hysteretic curve that displays an abrupt transition upon cooling through the MIT. N.B., the other transport properties of this sample are fully resembling a typical microscopic sample, including a high Figure of Merit current density. A false colored scanning electron micrograph of this sample is found in Supplementary Figure \ref{sub_fig_2}b. 

A possible origin of the observed broadening of the transition might be strain patterns in the sample caused by the coupling to the substrate. In the main text we argue that our microscopic samples do not experience any strain while driving a low current (current densities well below the IMT; including the FOM). Near the the IMT - independent whether it is induced by current or temperature - this no longer holds, as the lattice constants significantly alter and strain effect can dominate.\cite{Cirillo2019,Bertinshaw2019} Furthermore, a broadening of the transition is commonly observed in thin films, where strain caused by the substrate is a key parameter.\cite{Tsurumaki-Fukuchi2020,Miao2012,Wang2004}

 \begin{figure}[htb!]
 \centerline{$
 \begin{array}{c}
  \includegraphics[width=1\linewidth]{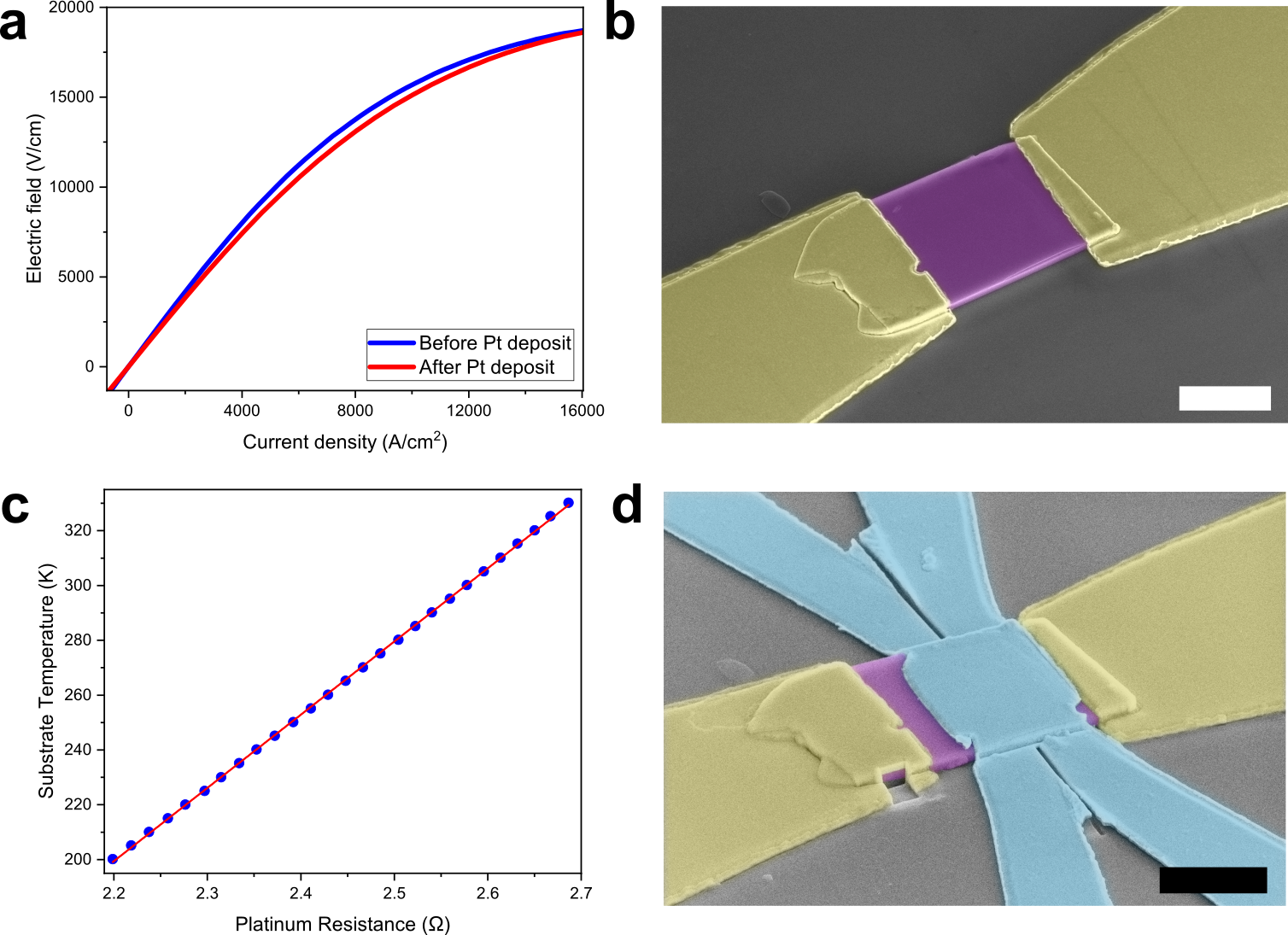}
 \end{array}$}
 \caption{\textbf{Images and transport properties of the embedded Pt-thermometer.} \textbf{a} Displays the $E(J)$-characteristic before and after the deposition of the Pt-thermometer. \textbf{b} Shows a false colored scanning electron microscope (SEM) image of the sample before the application of the thermometer and \textbf{d} shows a false colored SEM image that is acquired after the fabrication of the thermometer. In these images, the crystal is colored purple, the titanium/gold contacts to the crystal are yellow, and the platinum is light blue. The scale bar corresponds to 3 $\mu$m in both SEM images. In \textbf{c} the calibration curve of the Pt-thermometer, which is fitted by a linear curve, is shown. The slope of the fit corresponds to 0.27 K per m$\Omega$. This enables sub-Kelvin precision thermometry.}\label{sub_fig_1}
 \end{figure}
 
 \begin{figure}[htb!]
 \centerline{$
 \begin{array}{c}
  \includegraphics[width=1\linewidth]{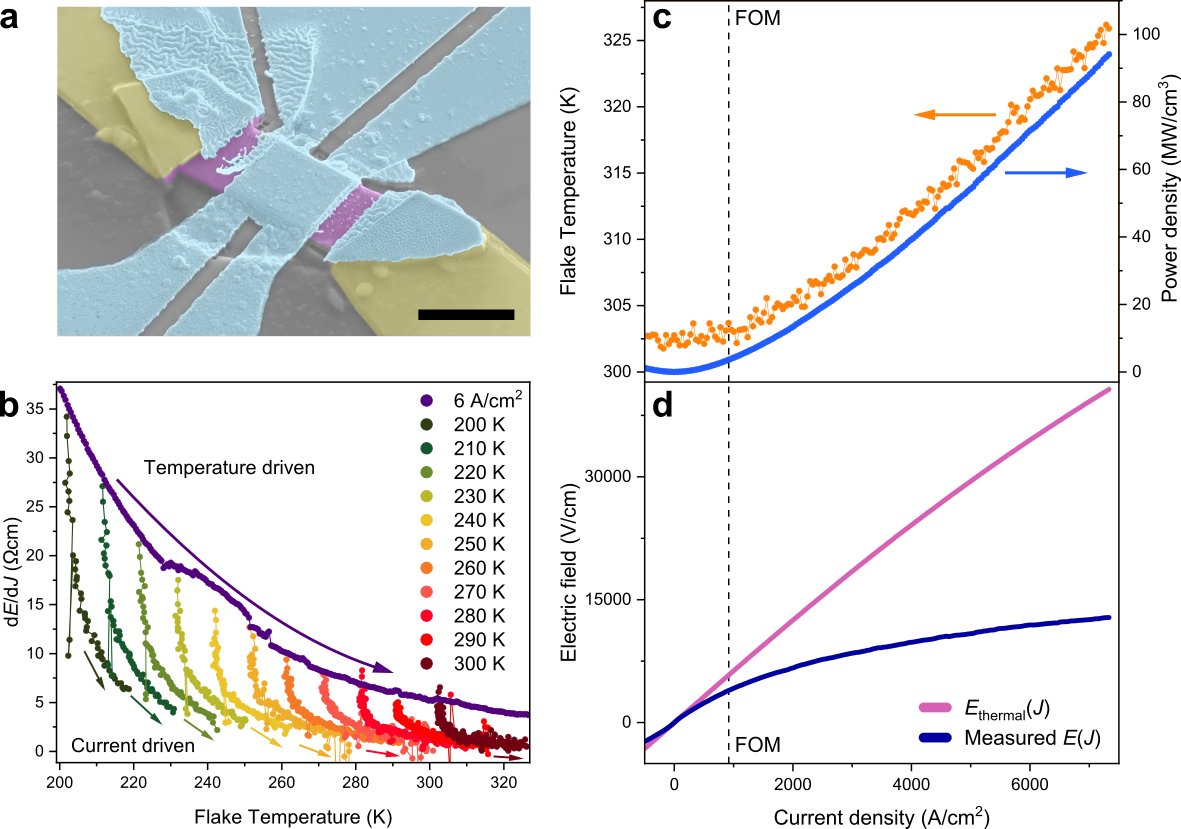}
 \end{array}$}
 \caption{\textbf{Complementary thermometry data set obtained on a microscopic sample.} \textbf{a} Shows a false colored scanning electron micrograph of a microscopic sample with a Pt-thermometer ($A = 1.8$ $\mu$m). The crystal is colored purple, the titanium/gold contacts are yellow and the platinum is indicated by light blue. The scale bar corresponds to 6 $\mu$m. In \textbf{b} the differential resistivity as a function of local sample temperature, analogous to Figure 3d of the main text, is shown. \textbf{c} depicts the local temperature and dissipated power density (product of $E$ and $J$) as a function of current density. A comparison between the measured $E(J)$-characteristic and $E_{\text{thermal}}$ (see section S5 and section 2.3 of the main text) is shown in \textbf{d}. The dashed line indicates the Figure of Merit current density.}\label{sub_fig_7}
 \end{figure}

 \begin{figure}[htb!]
 \centerline{$
 \begin{array}{c}
  \includegraphics[width=1\linewidth]{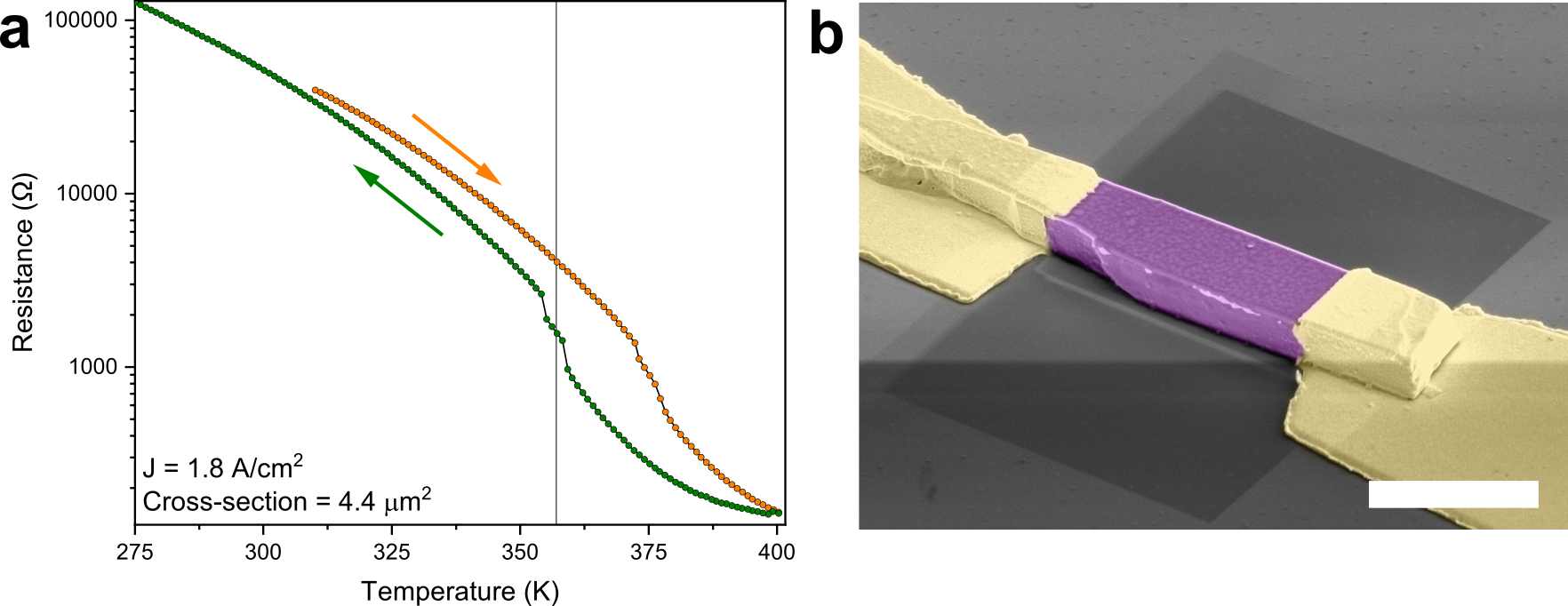}
 \end{array}$}
 \caption{\textbf{Resistance versus temperature of a relatively thick microscopic sample.} \textbf{a} Shows the resistance versus temperature of a microscopic sample that is 1.5 $\mu$m thick and 3 $\mu$m wide, which is relatively thick for a microscopic sample. We observe hysteresis on cycling the temperature and see an abrupt change in resistance when passing the thermal transition temperature of 357K as indicated by the vertical reference line. In \textbf{b} a false colored scanning electron micrograph is shown of the sample on which the data in \textbf{a} is acquired. Purple indicates the crystal and yellow colored are the titanium/gold contacts to it. The scale bar corresponds to 5 $\mu$m.}\label{sub_fig_2}
 \end{figure}

\clearpage

\subsection{The Figure of Merit as a function of other length scales}

In Figure 2b of the main text, we have chosen to plot the Figure of Merit as a function of the cross-sectional area. Here we present the same data as a function of thickness, width and length (between the voltage contacts) of the crystal, as plotted in Supplementary Figure \ref{sub_fig_3}. Most instructive is to inspect the samples on which we have performed a thinning study, these are highlighted by the use of a non-circular symbol in Supplementary Figure \ref{sub_fig_3}. If we first consider samples of constant thickness, we observe that the FOM can change an order of magnitude in a single sample while decreasing the width, which rules out a dependence solely on thickness. A similar argument can be made for the width and length dependence. Indeed, we have carried out a single step of decreasing the thickness of a sample, which is indicated by the use of arrows in Supplementary Figure \ref{sub_fig_3}. While the width is constant, the Figure of Merit changes again if we decrease the thickness of this sample. 

If we, however, plot the Figure of Merit as a function of the cross-sectional area of the sample, as is done in Figure 2b of the main text, we find that the thinning study samples follow a more coherent trend. We therefore use the cross-section as a typical length scale measure.

We can conclude from this observation that neither thickness, width or length of the crystal solely govern the Figure of Merit, meaning that none of these dimensions are more special in determining the transport properties of the crystal.

Supplementary Figure \ref{sub_fig_VOLUME_REB1} shows the FOM as a function of sample volume. Since $V^c = l^cA^c$ (where, $V$ is the volume, $l$ is the distance between the voltage contacts, $A$ is the sample cross-section, and $c$ is a constant) we see a very similar dependence of the FOM on volume as on the cross-sectional area.

\subsection{Additional information on FIB structuring}

The motivation for using focused ion beam (FIB) structuring in this work is two-fold. Employing FIB, we can structure the microscopic samples to have a rectangular cross-sectional area, which ensures full control over the current path. Furthermore, in order to rule out any geometry related artifacts, we performed a thinning study on three different samples corresponding to three different sample size ranges. By the use of FIB, we can sculpt the sample and reduce the cross-sectional area between consecutive $E(J)$-measurements. 

An often heard point of criticism on FIB processing is the implantation of Ga-ions in the material and in general damages of the crystal due to the high impulse of the ions. Key in resolving these issues is the combination of ultra-low beam currents and a SiO$_\text{x}$ capping layer. The first ensures that the damages are only very local (approximately within 30 nm from the edge of the milled structure), the second protects the top side of the crystal during inevitable radiation of Ga-ions.

In a previous work on the isostructural superconducting ruthenate Sr$_2$RuO$_4$, we have shown that it is possible to structure this complex oxides with up to 200 nm features without sacrificing sample quality.\cite{Yasui2020} Superconductivity in Sr$_2$RuO$_4$ is very sensitive to disorder and therefore producing samples with high residual resistivity ratios that show a literature value critical temperature, is a strong indication of the absence of any FIB induced damages. 
To illustrate the thinning experiment, Supplementary Figure \ref{sub_fig_5} shows false colored images of the three samples before thinning and after a couple of thinning steps.

 \begin{figure}[htb!]
 \centerline{$
 \begin{array}{c}
  \includegraphics[width=1\linewidth]{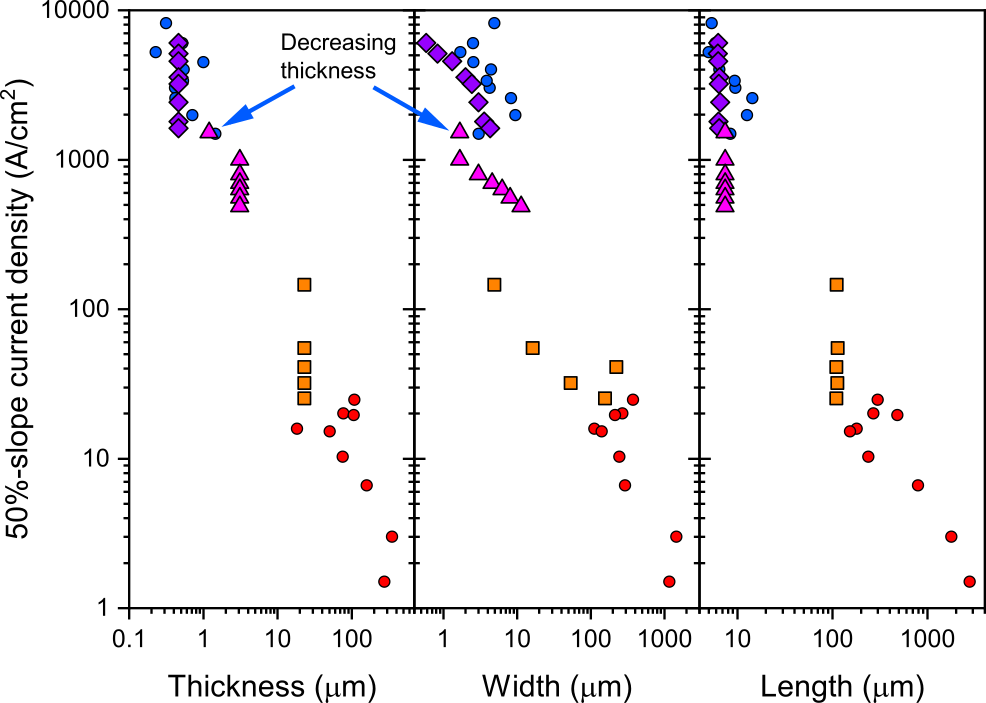}
 \end{array}$}
 \caption{\textbf{the 50\%-slope current density (FOM) shown as a function of different length scale parameters of the measured samples.} In this Figure, each circular symbol represents a single sample. The non-circular symbols represent the thinning study samples (see Supplementary Figure \ref{sub_fig_5} for images of these). The arrows indicate the decrease of thickness during the thinning study on the sample depicted in Supplementary Figure \ref{sub_fig_5}c and \ref{sub_fig_5}d.}\label{sub_fig_3}
 \end{figure}
 
 \begin{figure}[htb!]
 \centerline{$
 \begin{array}{c}
  \includegraphics[width=1\linewidth]{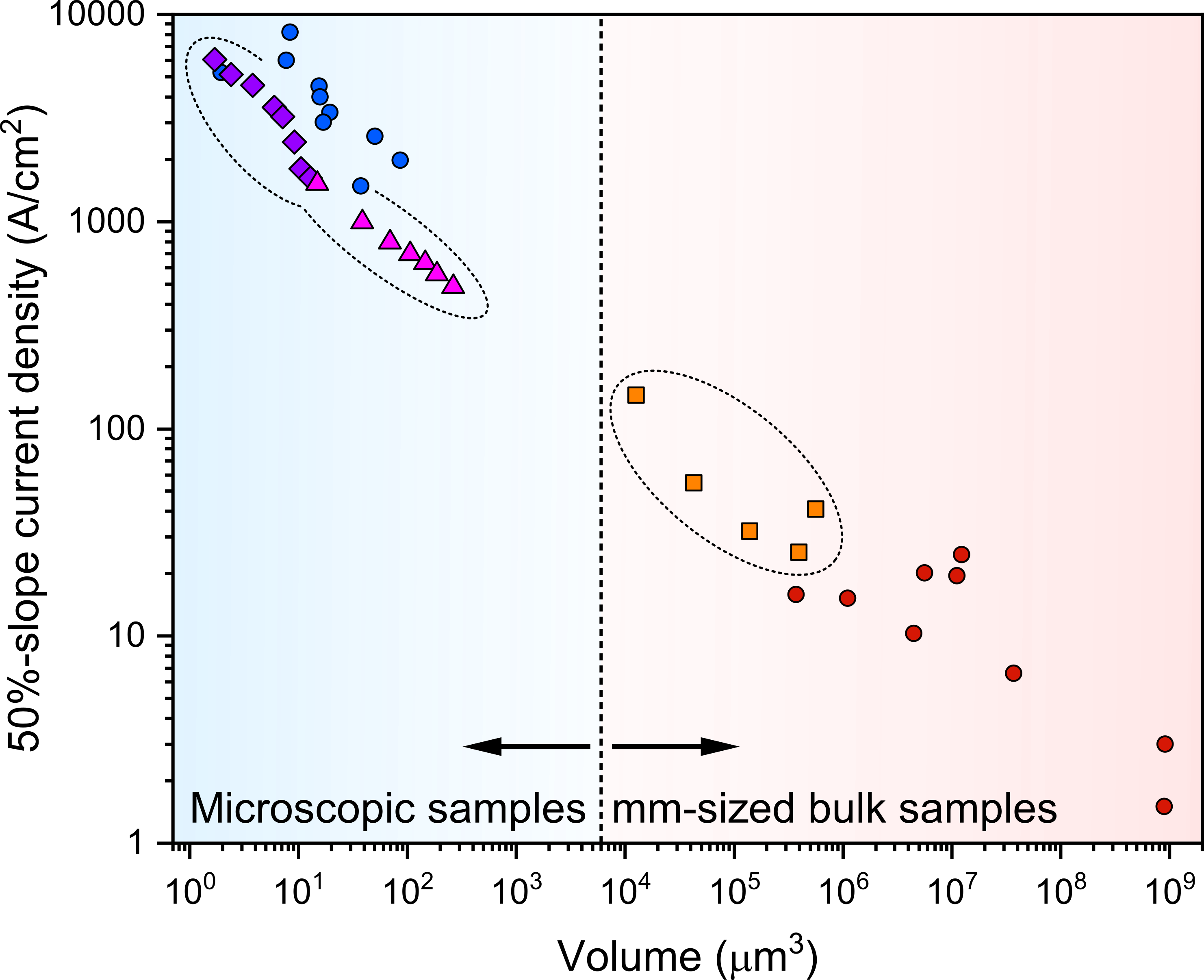}
 \end{array}$}
 \caption{\textbf{The Figure of Merit as a function of volume.}  The FOM as a function of volume for all measured samples at room temperature on a log-log scale. N.B. each circular point corresponds to a single sample. The thinning study samples are depicted with a unique non-circular symbol and a different color. The FOM shows a similar dependence on volume as on the cross-sectional area.}\label{sub_fig_VOLUME_REB1}
 \end{figure}

 \begin{figure}[htb!]
 \centerline{$
 \begin{array}{c}
  \includegraphics[width=1\linewidth]{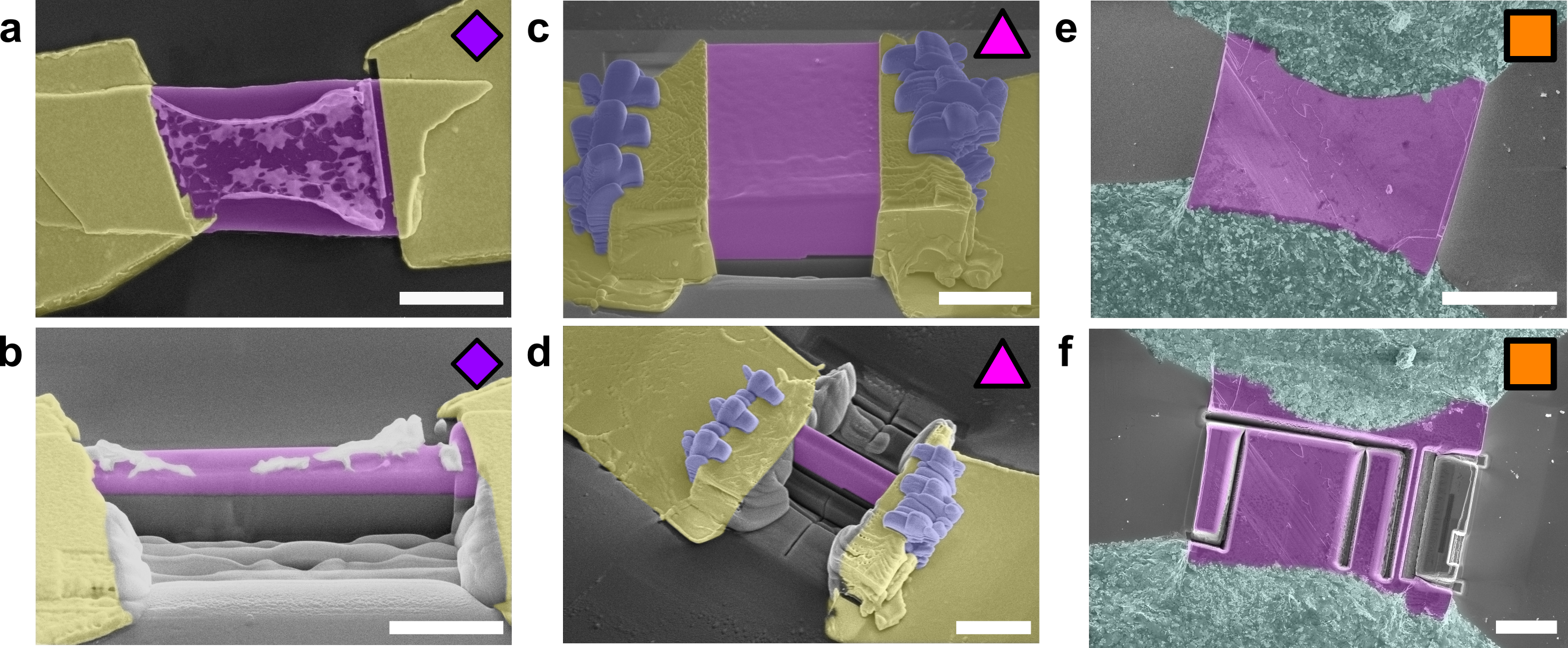}
 \end{array}$}
 \caption{\textbf{False colored scanning electron micrographs of samples used for the thinning experiment in different stages of thinning.} In all images, purple is used to indicate the crystal and yellow is indicating titanium/gold contacts to the sample. \textbf{a} and \textbf{b} correspond to the sample indicated by diamond signs in Figure 2b of the main text. It covers the smallest size ranges; the scale bars correspond to 3 $\mu$m and 2 $\mu$m respectively. Some lithography resist residue can be observed in white. The sample depicted in \textbf{c} and \textbf{d} is covering the intermediate sizes (scale bars are respectively 4 $\mu$m and 5 $\mu$m long) and is represented by triangles in Figure 2b of the main text. Indicated in blue in these images is the electron beam induced deposited tungsten used to strengthen the contacts to the crystal. The third sample of the thinning study is shown in images \textbf{e} and \textbf{f}. It is a relatively small mm-sized bulk sample that is contacted using silver epoxy, which is false colored in light blue. The FOM data of this sample is depicted by orange squares in Figure 2b in the main text. The scale bars in these images represent 100 $\mu$m and 50 $\mu$m respectively.}\label{sub_fig_5}
 \end{figure}
 
\clearpage

\subsection{Reconstructing the $E(J)$-characteristic on basis of Joule heating only}

Using analytical differentiation it is possible to calculate $\mathrm{d}E/\mathrm{d}J$ from $E(J)$. Reversely, one can calculate $E(J)$ by integration of $\mathrm{d}E/\mathrm{d}J$. The latter introduces an integration constant that needs to be set by a boundary condition. The boundary condition in our case is the requirement of zero electric field for zero current density: $E(0) = 0$. To summarize:

\begin{align} \label{Eq_sup:1}
E(J) \xrightarrow{\text{diff.}} \frac{\mathrm{d}E}{\mathrm{d}J}(J) \xrightarrow{\text{int.}} E_\text{recon.}(J) + C_\text{int.} \xrightarrow{\text{Boundary condition}} E_\text{recon.}(J) = E(J)
\end{align}

\noindent Here $C_\text{int.}$ is the integration constant and $E_\text{recon.}$ is the reconstructed electric field from the derivative $\mathrm{d}E/\mathrm{d}J$. As a sanity check, we perform differentiation and integration sequentially on acquired electric field and recover the originally measured data (see Supplementary Figure \ref{sub_fig_4}a). 

The resistivity as a function of temperature is measured by calculating the slope of the $E(J)$-characteristic for low currents (\textit{i.e.}, here, the slope of the $E(J)$ is current independent and no local temperature increase is measured). We can therefore interchange the resistivity as a function of temperature for the $\mathrm{d}E/\mathrm{d}J$ as a function of temperature at approximately zero applied current:

\begin{align} \label{Eq_sup:2}
\rho(T) = \frac{\mathrm{d}E}{\mathrm{d}J}(T, J \approx 0)
\end{align}

\noindent The application of the Pt-thermometer allows us to measure the local absolute temperature as a function of current density in the sample. When the crystal is heated, the resistivity lowers, causing a decrease of $\mathrm{d}E/\mathrm{d}J$ and therefore a decline of the slope of the $E(J)$-characteristic. Since we have both the $\mathrm{d}E/\mathrm{d}J$ as a function of temperature and the temperature as a function of current density available, we can combine these to find the $\mathrm{d}E/\mathrm{d}J$ as a function of the locally measured temperature caused by the application of current density, which we denote as $\mathrm{d}E_\text{thermal}/\mathrm{d}J$:

\begin{align} \label{Eq_sup:3}
\frac{\mathrm{d}E}{\mathrm{d}J}(T, J \approx 0) \And T(J)\xrightarrow{substitution} \mathrm{d}E_\text{thermal}/\mathrm{d}J(J)
\end{align}

\noindent To make this substitution, we fit the measured resistivity as a function of temperature with a 5th order polynomial to capture its phenomenological behavior (see Supplementary Figure \ref{sub_fig_4}b). Next, we use the fitted function to calculate $\mathrm{d}E_\text{thermal}/\mathrm{d}J$, which is depicted in Supplementary Figure \ref{sub_fig_4}c. Finally, we apply the above described procedure to reconstruct $E_\text{thermal}$ using integration, yielding the data that is shown in Figure 3 of the main text and Supplementary Figure \ref{sub_fig_4}d. 

 \begin{figure}[htb!]
 \centerline{$
 \begin{array}{c}
  \includegraphics[width=1\linewidth]{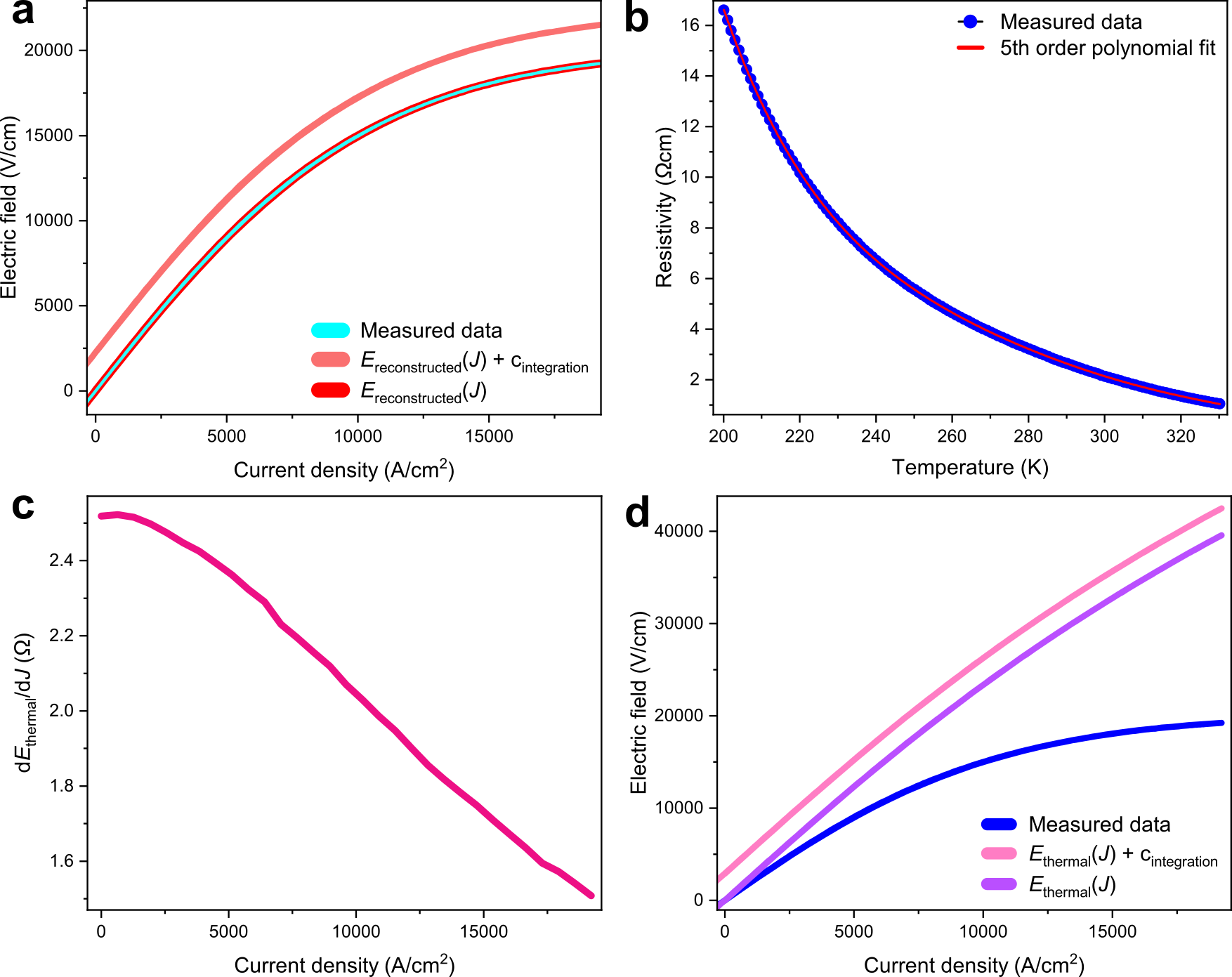}
 \end{array}$}
 \caption{\textbf{Summary of calculating $E_\text{thermal}$.} As a sanity check, we perform subsequent differentiation and integration of a dataset and retrieve the original measured data, which is depicted in \textbf{a}. \textbf{b} shows the resistivity versus temperature curve measured on a sample with embedded Pt-thermometer. The data is fitted by a 5th order polynomial to capture its phenomenological behavior. This fit is used to find the $\mathrm{d}E_\text{thermal}/\mathrm{d}J(J)$ (shown in \textbf{c}), which is integrated to retrieve $E_\text{thermal}$ (shown in \textbf{d} and Figure 3b of the main text).}\label{sub_fig_4}
 \end{figure}

\subsection{Details of the Comsol simulations and the homogeneity of the local temperature}

Since the resistance of the platinum thermometer is determined by its average temperature, our thermometry technique might be insensitive to local temperature inhomogeneity in the sample. Furthermore, due to the thermal coupling of the platinum leads to the substrate, a temperature difference might exist between the Pt and the part of the crystal that is most effected by heating. In order to eliminate these concerns, we have carried out thermal simulations in Comsol multiphysics version 5.4.

We consider the geometry displayed in Supplementary Figure \ref{sub_fig_6}b, which is a true to size model of the sample shown in Figure 1a of the main text. First, we assume that power is only dissipated in the part of the crystal between the gold contacts (\textit{i.e.}, not in the parts covered by gold). The input power is calculated from the product of current and voltage, taken from transport measurements (see Supplementary Figure \ref{sub_fig_6}a). Next, we assume that the substrate temperature is constant 50 $\mu$m away from the crystal. Using these two boundary conditions, we solve for a steady state solution where the temperature of the geometry is constant. The results of these simulations are governed by the thermal conductivity of the materials. Terasaki et al. have evaluated the in- and out-of-plane thermal conductivity of Ca214 to be 5.1 W/mK and 1.8 W/mK respectively.\cite{Terasaki2020}

In the main text, we report that the Pt-sensor is an accurate measure for the average sample temperature. Supplementary Figure \ref{sub_fig_6}c and \ref{sub_fig_6}d give an overview of the temperature variations within the sample at the FOM current density. Here, the temperature increase is plotted as a function of a vertical and horizontal line cut respectively. The temperature increases roughly quadratically from the sides of the crystal towards the hottest point of the flake. The temperature variations in the flake were found to be lower than 3 K at the FOM.

 \begin{figure}[htb!]
 \centerline{$
 \begin{array}{c}
  \includegraphics[width=1\linewidth]{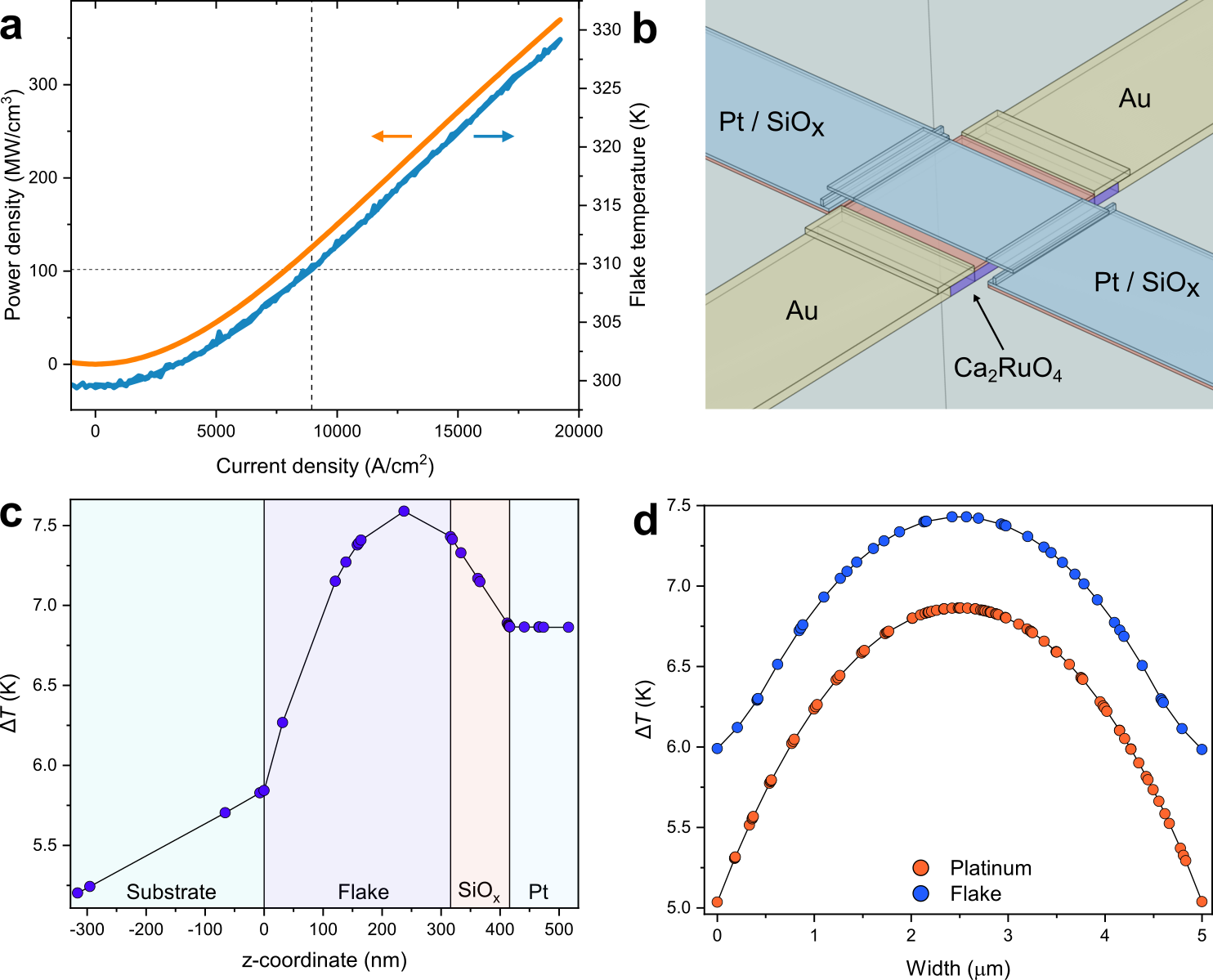}
 \end{array}$}
 \caption{\textbf{Temperature inhomogeneity in the Comsol simulations.} \textbf{a} displays the power density calculated from the transport data together with the measured local sample temperature. There is a one to one correspondence of the local temperature to the power density, giving a strong indication that we are measuring the local sample temperature. The dissipated power is used as input data for the simulations. \textbf{b} Shows a schematic overview of the different elements simulated. Each color corresponds to the labeled material.\textbf{c} depicts the temperature increase $\Delta T = T_{\text{flake}} - $300 K along a vertical line cut through the center of the substrate, flake and thermometer, at the FOM current density. The different parts of the sample are labeled by the same colors as in \textbf{b}. The temperature variations along a horizontal line cut parallel to the Pt-thermometer leads, at the FOM current density, are displayed in \textbf{d}.}\label{sub_fig_6}
 \end{figure}

\clearpage

\subsection{Minimal model for current density inhomogeneity over the cross-sectional area}

As described in the main text, the cross section dependence might be explained by an inhomogenous current density distribution throughout the cross-sectional area. Here we present a minimal toy model, and phenomenologically compare it to the measured data.

Crucial in this analysis is the difference between the \textit{apparent} current density, namely the total current (I) divided by the total cross-sectional area (A):

\begin{align} \label{Eq_sup_M1}
J_\text{app} = I / A
\end{align}

\noindent and the \textit{actual} current density that might be dependent on parameters describing the cross-sectional area:

\begin{align} \label{Eq_sup_M2}
J_\text{act} = \frac{\mathrm{d}I}{\mathrm{d}A} (t,w)
\end{align}

\noindent Where $t$ ($w$) is a length parameter that runs between $-T/2$ ($-W/2$) and $T/2$ ($W/2$), with $T$ ($W$) the thickness (width) of the crystal. The actual and apparent current density can be related to each other by integrating the actual current density to find the total current:

\begin{align} \label{Eq_sup_M3}
J_\text{app.} = I / A = 1/A \iint \frac{\mathrm{d}I}{\mathrm{d}A} (t,w) \mathrm{d}A
\end{align}

\noindent Therefore a homogeneous current distribution (\textit{i.e.}, a constant actual current density) yields an equal apparent current density:

\begin{align} \label{Eq_sup_M4}
J_\text{act.} = J_0 \xrightarrow{} J_\text{app} = 1/A  \iint J_0 \mathrm{d}A = \frac{A J_0}{A} = J_0 = J_\text{act.}
\end{align}

\noindent Now we can examine what type of actual current density features the power law dependence on the cross-sectional area that is observed in the measured current density (displayed in Figure 2b of the main text).
To do so we assume that the actual current density decreases as  $l^{-a}$, where $a$ is a constant and $l$ is the distance from the edges of the sample. This means, for a rectangular cross-section, that the total current is given by:

\begin{align} \label{Eq_sup_M5}
I = 4 \int_{0}^{T/2} \int_{0}^{W/2} \frac{\mathrm{d}I}{\mathrm{d}A}(t,w) \mathrm{d}t\mathrm{d}w
\end{align}

\noindent Here we used the fact that the system is symmetric in four sectors. Each of those can be divided in two parts:

\begin{align} \label{Eq_sup_M6}
I = 4 \left[ \int_{0}^{T/2} \int_{0}^{\frac{W}{T} t} w^{-a} \mathrm{d}w\mathrm{d}t + \int_{0}^{w/2} \int_{0}^{\frac{T}{W} w} t^{-a} \mathrm{d}t\mathrm{d}w \right]
\end{align}

\noindent Leading to:

\begin{align} \label{Eq_sup_M6}
I = \frac{4}{(1-a)(2-a)} \left[ \left(\frac{T}{W}\right)^{1-a} \left(\frac{W}{2}\right)^{2-a} + \left(\frac{W}{T}\right)^{1-a} \left(\frac{T}{2}\right)^{2-a} \right]
\end{align}

\noindent Where we assumed $a \neq 1$. This can be simplified to:

\begin{align} \label{Eq_sup_M7}
I = \frac{8}{(1-a)(2-a)} (TW)^{1-a} = \frac{8}{(1-a)(2-a)} A^{1-a}
\end{align}

\noindent That entails for the apparent current density:

\begin{align} \label{Eq_sup_M8}
J_\text{app.} = I / A = \frac{8}{(1-a)(2-a)} \frac{A^{1-a}}{A} = \frac{8}{(1-a)(2-a)} A^{-a} \sim A^{-a}
\end{align}

\noindent Which is the observed power law dependence of the measured current density.